\documentclass[12pt,aps,prx,nofootinbib]{revtex4-2}
\usepackage[utf8]{inputenc}
\usepackage{amsmath}
\usepackage{amsfonts}
\usepackage{amsthm}
\usepackage{amssymb}
\usepackage{graphicx}
\usepackage[usenames,dvipsnames]{color}
\usepackage{hyperref}
\usepackage{mathrsfs}

\theoremstyle{definition}

\theoremstyle{remark}

\newcommand{\abs}[1]{\left\vert#1\right\vert}
\newcommand{\set}[1]{\left\{#1\right\}}

\newcommand{\expec}[1]{\left\langle#1\right\rangle}

\newcommand{\ket}[1]{\left|#1\right\rangle}

\newcommand{\lag}{\mathscr{L}}

\def\actaa{\ref@jnl{Acta Astron.}}      

\def\prd{\textrm{Phys.~Rev.~D}}        
\def\physrep{\textrm{Phys.~Rep.}}   

\begin{document}

\title{The Equivalence Principle and The Cosmological Constant Problem\footnote{Essay written for the Gravity Research Foundation 2021 Awards for Essays on Gravitation}}%
\author{Yonadav Barry Ginat}%
\affiliation{Faculty of Physics, Technion -- Israel Institute of Technology, Haifa, 3200003, Israel}%
\email{ginat@campus.technion.ac.il}%

\begin{abstract}
  In this essay I point out that, in the context of semi-classical gravity, the equivalence principle can mitigate the cosmological constant problem. On a Minkowski space-time background with the usual $\mathbb{R}^4$ topology, the vacuum self-energy is removed by normal ordering; this is allowed because it is not observable; I argue that, in a freely-falling frame of reference, the same must hold true, up to contributions from modes whose wavelength is of the order of the background radius of curvature. Thus, the equivalence principle implies that ultra-violet modes do not contribute to the effective energy-momentum tensor.
\end{abstract}

\maketitle

\newpage

The cosmological constant problem is the discrepancy between the small value of the cosmological constant, inferred from observations ($\sim 10^{-47}~ \textrm{GeV}^{4}$ \cite{Planck2018}), and its expected theoretical value, estimated from dimensional analysis, to be of order $l_p^{-4} \sim 10^{71} ~\textrm{GeV}^{4}$ \cite{Zeldovich1968}. Many ideas were proposed to address this problem (see, e.g., \cite{Weinberg1989,Polchinski2006,Bousso2012} and references therein), many focused on extensions of general relativity. Short of a full theory of quantum gravity, I also endeavour to alleviate it, but without changing general relativity or quantum field theory -- rather, by performing a careful semi-classical analysis. Specifically, I wish to explain here why ultra-violet contributions to bubble diagrams (diagrams without external vertices) in the vacuum-to-vacuum amplitude do not gravitate, even when the underlying gravitational field is dynamical.

One can also look at the problem from another angle: the renormalised vacuum expectation value of the energy-momentum tensor, $\expec{0|T_{ab}|0}_{\rm ren}$, is uniquely determined by certain plausible axioms, together with its value for a non-dynamical Minkowski background \cite{Wald1994}; the latter is of course $k \eta_{ab}$ (which would become a term $kg_{ab}$ in $\expec{0|T_{ab}|0}_{\rm ren}$ for a general background), where $k$ is some constant, usually chosen as $0$. The cosmological constant problem is, essentially, the problem of justifying choosing $k=0$ when the metric is elevated from a non-dynamical background to a dynamical field, and not some other finite value, and the argument from the equivalence principle presented below shows why this must be so.

In semi-classical gravity, the equations of motion of the (classical) metric are derived from an effective action, related to the matter path integral, $Z_m[g]$. For simplicity, consider a free scalar field with mass $m$, governed by
\begin{equation}
  Z_m[g_{ab}] = \frac{1}{N[g_{ab}]}\int \mathcal{D}\varphi~\exp\left[-\frac{\mathrm{i}}{2\hbar}\int_{\mathcal{M}}\mathrm{d}^4x \sqrt{\abs{g}} \left(\partial_a \varphi \partial^a \varphi + \frac{m^2}{\hbar^2}\varphi^2\right)\right] \equiv \int \mathcal{D}\varphi \frac{e^{\mathrm{i}S_m[\varphi]/\hbar}}{N[g_{ab}]},
\end{equation}
where $g = \det g_{ab}$, $\mathcal{M}$ is space-time, and $N \propto \sqrt{\det G_{\rm ret}}$ \cite{DeWitt1975}, where $G_{\rm ret}$ is the retarded Green's function of $\varphi$.

The most appropriate manner to obtain the correct action for $g_{ab}$ is by integrating the matter degrees of freedom out, yielding an effective action for gravity, $\Gamma[g_{ab}]$, that should contain the gravitational effect of $\varphi$. In practice it is impossible to do so precisely, since one would be required to find a general functional form of the Green's function of $\varphi$, as a function of the metric $g_{ab}$. As one intends to vary the effective action with the metric, it is important to treat it as completely general. Therefore, one has to recourse to approximations in integrating out the matter fields; the way to do so is by invoking the equivalence principle.

This is done as follows: assume that the radius of curvature $r_c$ of $({\mathcal{M}},g_{ab})$, is considerably larger than the characteristic length-scales of the matter action, namely
\begin{equation}
  r_c \gg r_m \equiv \max\set{\frac{\hbar}{m},Gm},
\end{equation}
(this is of course true also for all standard model fermions). If so, then the propagator of $\varphi$ is dominated by scales of order $r_m$, to leading order.
Then, by the equivalence principle, any local experiment cannot tell the difference between a gravitational field and a non-inertial frame of reference, over the relevant length scales of the standard model. In quantum physics, observables are determined by the path integral -- therefore it is the effective action to which the equivalence principle directly applies, not the action. Let $Q_{\mathbf{a}}(x)$ be some local tensor operator, corresponding to some locally measurable quantity,\footnote{`Local' in the sense of the equivalence principle, i.e. whose support about some point $x$ is much smaller than $r_c$. $Q_{\mathbf{a}}$ can also depend on more than one point, provided that their distance is $\ll r_c$.} where $\mathbf{a} = a_1\ldots a_n$ is a multi-index. Its expectation value in flat space-time is
\begin{equation}\label{eqn:local expectation value}
  \langle Q_{\mathbf{a}}^{\rm flat} \rangle = \lim_{\lambda \to \infty}\frac{1}{N}\int \mathcal{D}\varphi~ Q_{\mathbf{a}}[\varphi]e^{\mathrm{i}S_m/\hbar}
\end{equation}
where $\lambda$ is a collective name for a regulator in some regularisation scheme, which applies to both $N$ and the matter path integral. Crucially, $\langle Q_{\mathbf{a}}^{\rm flat}(x) \rangle$ is renormalised locally, for otherwise it would not satisfy the equivalence principle (ref. \cite{Shapiro_2008} shows that counter-terms are indeed local). Since $Q_{\mathbf{a}}^{\rm flat}$ is a Lorentz tensor in Minkowski space-time, the principle of general covariance of the laws of physics dictates that it can be promoted to a tensor $\langle Q_{\mathbf{a}}^{\mathcal{M}}(x)\rangle$ on $\mathcal{M}$, which is given by equation \eqref{eqn:local expectation value} in Riemann normal co-ordinates (RNCs) about $x$.

What happens to bubble diagrams? They are automatically cancelled by $N$, and hence cannot contribute to $\langle Q_{\mathbf{a}}^{\rm flat} \rangle$ at all. In other words, to compute $\langle Q_{\mathbf{a}}^{\mathcal{M}}(x)\rangle$, one has to transform to RNCs centred at $x$, compute $\langle Q_{\mathbf{a}}^{\rm flat} \rangle$, and then to transform back. Nothing in this process involves bubble diagrams, and therefore they do not contribute to $\langle Q_{\mathbf{a}}^{\mathcal{M}}(x)\rangle$ in any way.
(Here I assume for simplicity that the vacuum $\ket{0}(x)$ of the flat QFT is the same, in RNCs, for all $x$, and that $\expec{\varphi}$ vanishes.) The question one needs to answer is whether $\expec{0|0}$, given a certain metric $g_{ab}$, depends on $g_{ab}$. The equivalence principle implies that it does not (up to contributions from long modes, discussed below), because it is a \emph{local} quantity, and locally no measurement result (i.e. amplitude) is affected by it.

Working in RNCs amounts to an approximation of the integral $\int \mathrm{d}^4 x$ in $S_m$ by an integral over a region of size $R$, where $r_m \ll R \ll r_c$ -- an approximation which is justified by locality.
Contributions from long modes, i.e. modes whose wavelength is $\sim r_c$, only yield an $O(r_c^{-4}\hbar)$ correction to the effective action (this can follow either from a steepest descents-type approximation for the matter path integral, of simply from dimensional analysis). Observe, that the above arguments, based on the equivalence principle, imply that it is only potentially the long modes that contribute to a non-vanishing $\expec{T_{ab}}_{\rm ren}$, which is consistent with, e.g., ref. \cite{SciamaCandelasDeutsch1981}, which set $k=0$. Thus, $\Gamma[g_{ab}]$ contains no bubble diagram contributions, to an accuracy of $O(r_c^{-4}\hbar)$, i.e.
\begin{equation}
  \lag_{\rm eff} = \sqrt{-\det g}\left[\frac{R-2\Lambda}{16\pi G} + O(\hbar r_c^{-4})\right].
\end{equation}
This agrees with known results for $\Gamma$, obtained by adiabatic regularisation, which implicitly assumed that $k=0$ \cite{parker_toms_2009}. A discussion of possible infra-red divergences \cite{hu_verdaguer_2020} is beyond the scope of this essay, and does not occur for a massive field, anyway. Within the framework of semi-classical gravity, the $O(\hbar r_c^{-4})$ correction should be treated perturbatively \cite{ParkerSimon1993}, as it is proportional to $\hbar$. I neglect this correction in this essay, since it does not relate directly to the problem at hand.

In a non-dynamical Minkowski background, the 1-loop effective potential takes the form\footnote{Note that it is dominated by UV modes, whose wavelengths are $\lesssim r_m$.}
\begin{equation}
  V_{\rm eff} \sim \frac{-\mathrm{i}\hbar}{2}\int \frac{\mathrm{d}^4k}{(2\pi)^4}\ln\left[\frac{\mathrm{i}}{\pi}(k^2 + \frac{m^2}{\hbar^2})\right] = \frac{m^4\ln m^2}{64\pi^2\hbar^3} + (\mbox{infinite}) + (\mbox{infinite})m^2 + (\mbox{infinite})m^4,
\end{equation}
which is na\"{i}vely renormalised to $k = \frac{m^4\ln m^2}{64\pi^2}$, so that the path integral $Z_m = \sqrt{\det G_F}$, where $G_F$ is the Feynman propagator, becomes
\begin{equation}
  Z_m = \sqrt{\det G_F} = \exp\left(\frac{\mathrm{i}}{\hbar}\int \mathrm{d}^4x\sqrt{\abs{g}}V_{\rm eff}\right).
\end{equation}
The argument above implies that this contribution to the effective potential is cancelled by $N$, or, equivalently, by the measure of the gravitational field path integral; i.e., if $V_{\rm eff}$ is not regularised to zero, then $\sqrt{\det G_{\rm ret}}$ in $N$ receives an exactly equal contribution, which cancels it, ensuring $k=0$ (again, up to terms of higher order in $r_c^{-1}$). This is equivalent to using local normal ordering in these determinants, to compute the vacuum energy.
In-so-far-as the effective action $\Gamma[g_{ab}]$ is concerned, no local observables could ever be influenced by bubble diagrams, and therefore it cannot contain a contribution where such diagrams can couple to the metric. This is akin to saying that in computing it, the equivalence principle requires that the vacuum energy contribution to it be renormalised to zero locally, at any point in space-time, i.e. that when writing
\begin{equation}
  \frac{\mathrm{i}\Gamma[g_{ab}]}{\hbar} = \lim_{\lambda \to \infty} \ln \left[\frac{Z_m[J,g_{ab}]}{N[g_{ab}]}\right] + \frac{\mathrm{i}S_{EH}}{\hbar},
\end{equation}
where $S_{EH}$ is the Einstein-Hilbert action, for some source $J$ (e.g. a hydrogen atom), $N$ is understood to remove all bubble diagrams from the numerator, and both $Z_m$, $S_{EH}$ and $N$ contain all the necessary counter-terms. One immediate consequence of this, e.g. for a hydrogen atom background, is that loop corrections to its binding energy do gravitate, but loop diagrams in the vacuum do not.
(The dynamics of the gravitational field are of course dictated by $\Gamma$ plus the Einstein-Hilbert action, plus any action for classical matter that might be present.)

Let me also mention that this approximation is self-consistent: had one not been careful to regularise $\Gamma[g_{ab}]$ locally, the energy-momentum tensor induced by it would have contained a vacuum energy term $\sim \frac{r_m^{-4}}{G}g_{ab}$, which would have induced a huge back-reaction on $g_{ab}$, and would have changed $r_c$ to roughly $r_m$. This would have resulted in the invalidation of the approximation $r_c \gg r_m$, and thus would have rendered the above treatment inconsistent. However, since it has been shown above that the equivalence principle requires that vacuum diagrams not gravitate, and therefore that they should be locally regularised to zero in $\Gamma$, the remaining correction to the classical action is actually small. The local approximation in this essay and its consistency are thus saved, and the link between the vacuum energy of $\varphi$ and $\Lambda$ -- severed. Although in this essay I assumed that matter was composed of a free scalar field, the argument presented here can be straightforwardly generalised to general matter fields, provided they are minimally coupled, as well as to cases where spontaneous symmetry breaking occurs, where the correct value of $k$ would then correspond to the value of the potential at the true vacuum.

\acknowledgments{I am grateful to Vincent Desjacques, Hagai Perets and Oren Bergman for helpful discussions. I acknowledge support from the Israeli Academy of Sciences' Adams Fellowship.}





\bibliography{lambda_bibliography}

\end{document}